\documentclass{aa}
\usepackage[varg]{txfonts}
\usepackage{graphicx}
\usepackage{xcolor}
\bibpunct{(}{)}{;}{a}{}{,} 

\def\cm{cm$^{-1}$}

\begin{document}

\title{Stellar population astrophysics (SPA)
with the TNG\thanks{Based on observations made with the Italian Telescopio Nazionale Galileo (TNG)
operated on the island of La Palma by the Fundación Galileo Galilei of the INAF (Istituto Nazionale di Astrofisica) at the Spanish Observatorio
del Roque de los Muchachos of the Instituto de Astrofisica de Canarias.
This study is part of the
Large Program titled {\it SPA - Stellar Population Astrophysics:  the detailed, age-resolved
chemistry of the Milky Way disk} (PI: L. Origlia), granted observing time with HARPS-N and GIANO-B echelle spectrographs
at the TNG.}}
\subtitle{Identification of  a Sulphur line at  $\lambda_\mathrm{air}=1063.600$\,nm in GIANO-B stellar spectra}
   \author{N. Ryde \inst{1}
   \and 
  H. Hartman\inst{2,1}
            \and 
          E. Oliva\inst{3}
            \and
          L. Origlia\inst{4}
             \and
          N. Sanna\inst{3}
            \and
          M. Rainer\inst{3}
             \and
         B. Thorsbro\inst{1}   
           \and
           E. Dalessandro\inst{4}
           \and 
           G. Bono\inst{5,6}
           }
   \institute{Lund Observatory, Department of Astronomy and Theoretical Physics, Lund University, Box 43, SE-221 00 Lund, Sweden\\
             \email{nils.ryde@astro.lu.se}
          \and
             Materials Science and Applied Mathematics, Malm\"o University, SE-205 06 Malm\"o, Sweden
            \and
              INAF-Arcetri Astrophysical Observatory, 
              Largo E. Fermi 5, I-50125 Firenze (Italy)\\
               \email{oliva@arcetri.astro.it} 
            \and
              INAF - Osservatorio di Astrofisica e Scienza dello Spazio di Bologna,
             Via Gobetti 93/3, I-40129, Bologna, (Italy)
             \and 
             Dipartimento di Fisica e Astronomia, Universit\'a degli Studi di Roma Tor Vergata,
             Via della Ricerca Scientifica 1, I-00133, Roma (Italy)
        \and
            INAF - Osservatorio Astronomico di Roma,
            Via Frascati 33, I-00040, Monte Porzio Catone (Italy)
                  }
   \date{Received Month XX, 2019; accepted Month XX, 2019}

 
  \abstract
   {In the advent of new infrared,  high-resolution spectrometers, accurate and precise atomic data  in the infrared is urgently needed. Identifications, wavelengths, strengths, broadening and hyper-fine splitting parameters of stellar lines in the near-IR are in many cases not accurate enough to model observed spectra, and in other cases even non existing. Some stellar features are unidentified.}
   {The aim with this work is to identify a spectral feature at  $\lambda_\mathrm{vac}=1063.891$\,nm or $\lambda_\mathrm{air}=1063.600$\,nm seen in spectra of stars of different spectral types, observed with the GIANO-B spectrometer.}
   {Searching for spectral lines to match the unidentified  feature in linelists from standard atomic databases was not successful.  However, by investigating the original, published laboratory data we were  able to identify the feature and solve the problem. To confirm its identification, we model the presumed stellar line in the solar intensity spectrum and find an excellent match.  
   }
   {We find that the observed spectral feature is a stellar line originating from the 4s'--4p' transition in \ion{S}{I}, and that the reason for its absence in atomic line databases is a neglected air-to-vacuum correction in the original laboratory measurements from 1967 for this line only. From interpolation we determine the laboratory wavelength of the \ion{S}{I} line to be $\lambda_\mathrm{vac}= 1063.8908$\,nm or $\lambda_\mathrm{air}= 1063.5993$\,nm, and the excitation energy of the {\it upper} level to be 9.74978 eV.}
   {}


\maketitle
%

\section{Introduction}

In recent years several new cross-dispersed, near-infrared spectrometers
have been developed. These can capture large portions of one or more infrared bands (Y, J, H, K, L and/or M bands) 
simultaneously, which in principle increases the near-IR observing efficiency dramatically. 
Examples of high-resolution ($R>40,000$) spectrometers  are  GIANO-B \citep{giano14}, IGRINS \citep{igrins}, 
WINERED \citep{winered:18}, and CRIRES+ \citep{crires:14,crires:18} spectrometers. Also several medium-resolution ($R\sim 20-30,000$) spectrometers, such as APOGEE \citep{apogee:16,apogee:17}
and NIRSPEC \citep{nirspec_mclean}  can record large parts of a near-IR band in one setting. This will also be the case of the MOONS spectrometer \citep{moons:16,moons:18} which is under development and will be placed at the VLT.
Furthermore, the next generation of extremely large telescopes with their huge apertures will provide  enhanced sensitivity in medium-high resolution near-IR spectroscopy, thanks to spectrometers such as MOSAIC \citep[$R\sim 15,000$;][]{mosaic:18} and HIRES  \citep[$R\sim 100,000$; ][]{elt_hires:18} for the ELT.

The near-IR spectral region is therefore emerging as a spectral domain for versatile astrophysical use in an efficient way. 
However, since the near-IR wavelength region has not been as explored as the optical region for astrophysical use, the spectral data, such as line identifications,
wavelengths, strengths, broadening and hyper-fine splitting parameters,  are lagging behind. 
These data are, nevertheless, vitally needed for any spectral investigation, both at high and low spectral resolution \citep[see e.g.][]{ruffoni:15}. 

Experimental and theoretical progress is, however, being made to ameliorate the situation. Recent work include, for instance, 
experimental oscillator strengths of 28 Fe\,{\sc i} lines in the H band ($1.4-1.7\,\mu$m), following an urgent need from the APOGEE survey \citep{ruffoni:13} and
measurements of line strengths of
magnesium \citep{pehlivan:mg} and scandium \citep{pehlivan:sc}. These Sc data were vital in, for example, the discussion on K-band Sc abundances in the Nuclear Star Cluster by \citet{thorsbro:18}. 
Other examples are the works on WINERED spectra by \citet{winered_fe:19} who investigated Fe-lines at $0.9-1.3\,\mu$m in spectra of red giants and their usefulness for determining the metallicity and microturbulence and the work by \citet{fukue:15} concerning the development of the line-depth ratio method to determine effective temperatures of classical Cepheids in the H band. Furthermore, in the K band, not many measurements on relevant atomic data, such as wavelengths and lines strengths, from laboratory spectra exist at all and the line wavelengths in the VALD database \citep[e.g.][]{vald}  are often uncertain. An example taken from \citet{thorsbro:16} is a Si\,{\sc i} line at 2114.4\,nm with atomic physics parameters estimated theoretically \citep{kurucz:07}. This line can be identified in the solar spectrum to have a wavelength shift of as much as  0.1\,nm. Correcting the wavelength and checking against other stars, like Arcturus, shows that its identification is correct.  
\citet{thorsbro:16} further shows that there is a clear difference in the wavelength accuracy in the K band for different elements, sulphur having the smallest spread, and silicon and calcium requiring the largest corrections in order to match the solar spectrum.

\begin{figure*}[ht!]
\centering
\includegraphics[trim={1.5cm 4cm 0cm 4cm},clip,width=1.0\linewidth]{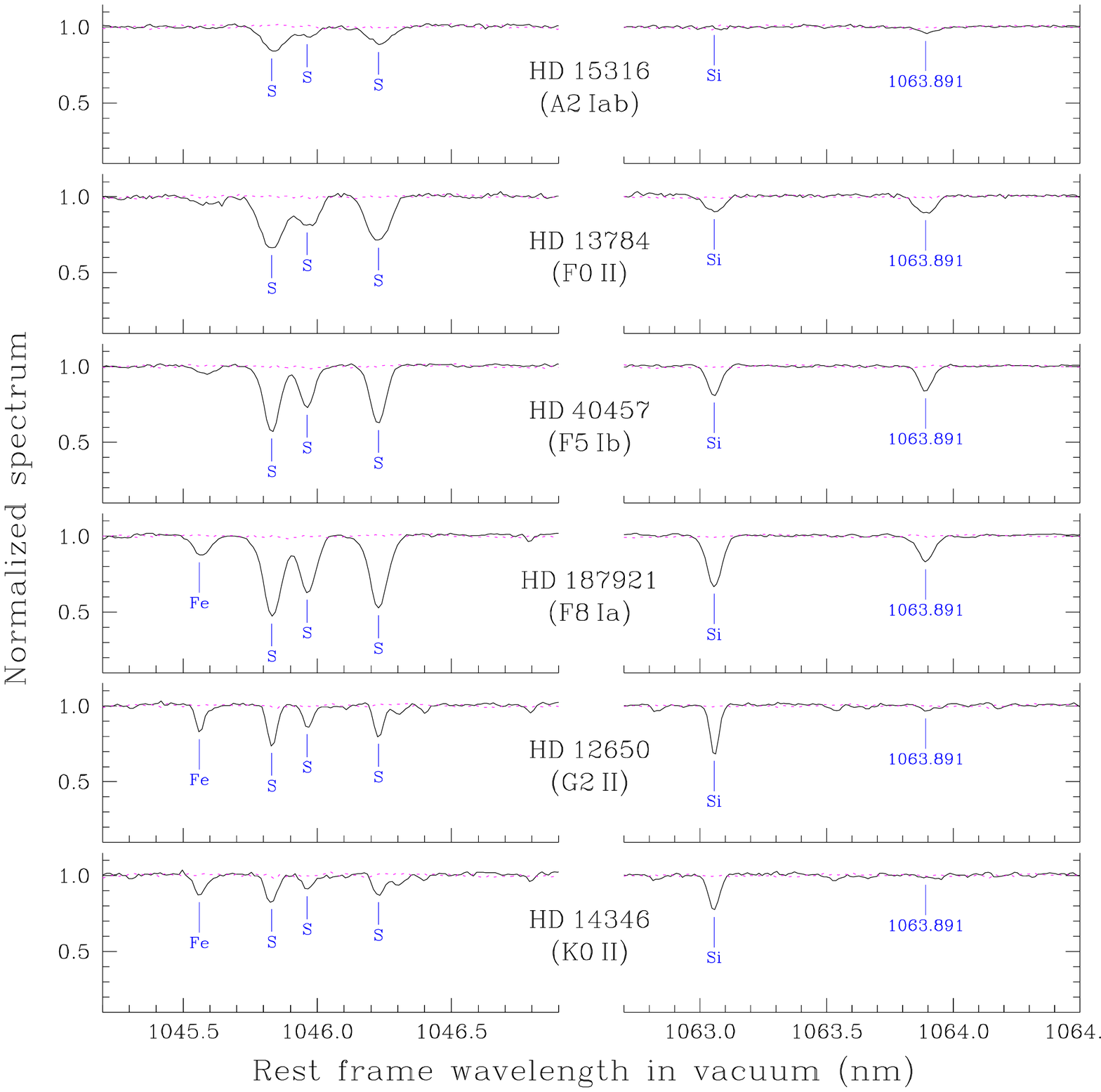}
\caption{Sections of the GIANO spectra including the unidentified feature at $\lambda_\mathrm{vacuum}=1063.891\,\mathrm{nm}$, identified as a Sulphur line in this paper. The sulphur triplet at 1046 nm are also shown.}
\label{fig:giano_spectrum}
\end{figure*}

\begin{table*}
\caption{Stellar Parameters}             
\label{param}      
\centering          
\begin{tabular}{l c c c c l }     
\hline\hline       
Name & Spectral Type &$T_\mathrm{eff}$ & $\log g$ & [Fe/H] & reference \\ 
\hline                    
  HD15316 & A2Iab & 8770   & 1.7  & -     & \citet{verdugo:99} \\
  HD13784 & F0II2 & 7080   & 2.1  &  0.0 & \citet{luck:14}  \\
  HD40457  (CO Aur) &  $\sim$F5Ib  (classical Cepheid)  & 6620   & 2.5  & -0.1 & \citet{luck:14}  \\
  HD187921  (SV Vul) &  $\sim$F8Ia (classical Cepheid) & 4900 - 6330 & 0.29 - 1.28 & 0.0 &\citet{luck:18}   \\
  HD12650 &  G2II  & 5325  & 2.7 & - & \citet{hohle:10}\\
  HD14346 & K0II & 4585 & 2.2 & - & \citet{hohle:10} \\
\hline                  
\end{tabular}
\tablefoot{The effective temperatures of the Cepheids  depend on the phase at which the spectrum was collected, the same applies for the spectral type.}
\end{table*}

In the same spirit, we present in this paper the identification of an unidentified feature at $\lambda_\mathrm{vac}=1063.891$\,nm in a range of stellar spectra observed at high spectral resolution with the GIANO-B spectrometer. We discuss the line's behaviour with stellar effective temperature and finally identify it as a S\,{\sc i} line. We also discuss why it was missed in earlier works.


\section{Observations and spectral analysis}

Six  giants and supergiants of spectral types from A to K were observed with GIANO-B, the high resolution (R$\simeq$50,000) infrared (950--2450 nm) spectrometer \citep{oli12a,oli12b,giano14} of the Telescopio Nazionale Galileo (TNG). The stars and their stellar parameters are provided in Table \ref{param}.

GIANO was designed for direct feeding of
light at a dedicated focus of the TNG. In 2012 the instrument was provisionally commissioned and used in the "GIANO-A" configuration; with the spectrometer positioned on the rotating building and fed via a pair of fibers connected to another focal station (\cite{tozzi14}). In 2016 the spectrometer was eventually moved to the originally foreseen configuration (called "GIANO-B") where it can also be used in the "GIARPS" mode for simultaneous observations with HARPS-N
\citep{tozzi16}.

GIANO provides a fully automated online data reduction pipeline based on the "GOFIO" reduction software (\cite{gofio}) that processes all the observed data; from the calibrations (darks, flats and U-Ne lamps taken in day-time) to the scientific frames. The main feature of the GOFIO data reduction is the optimal spectral extraction and wavelength calibration based on a physical model of the spectrometer that accurately matches instrumental effects such a variable slit tilt and orders curvature over the echellogram \citep{giano2Dreduction}. 

The spectra presented here were collected in November 2018 with the spectrometer in the "GIARPS" configuration. For an optimal subtraction
of the detector artifacts and background, the spectra were collected nodding the star along the slit; i.e. with the target alternatively positioned at 1/4 (position A) and 3/4 (position B) of the slit length.
Integration time was 5 minutes per A,B position. The nodding sequences were repeated to achieve a total integration time between 40 and 60 minutes per target.

The telluric absorption features were corrected using the spectra of a telluric standard (O-type star) taken at different airmasses during the same nights. The normalized spectra of the telluric standard taken at low and high airmass values were combined with different weights to match the depth of the telluric lines in the stellar spectra. 

Figure \ref{fig:giano_spectrum} 
shows the normalized GIANO-B spectra of stars with different temperatures.
The telluric correction applied is shown as a dashed line; it is negligible in the spectral region of interest here.
The positions of the stronger atomic lines are marked.
The wavelength scale is in the rest frame of each star, determined using standard cross-correlation techniques including the full GIANO-B spectrum (970 - 2450 nm). The final accuracy in the spectral region of interest here is 0.001 nm  rms.


\begin{figure*}[t]
\centering
\includegraphics[trim={0cm 0cm 0cm 0cm},clip,width=0.48\linewidth]{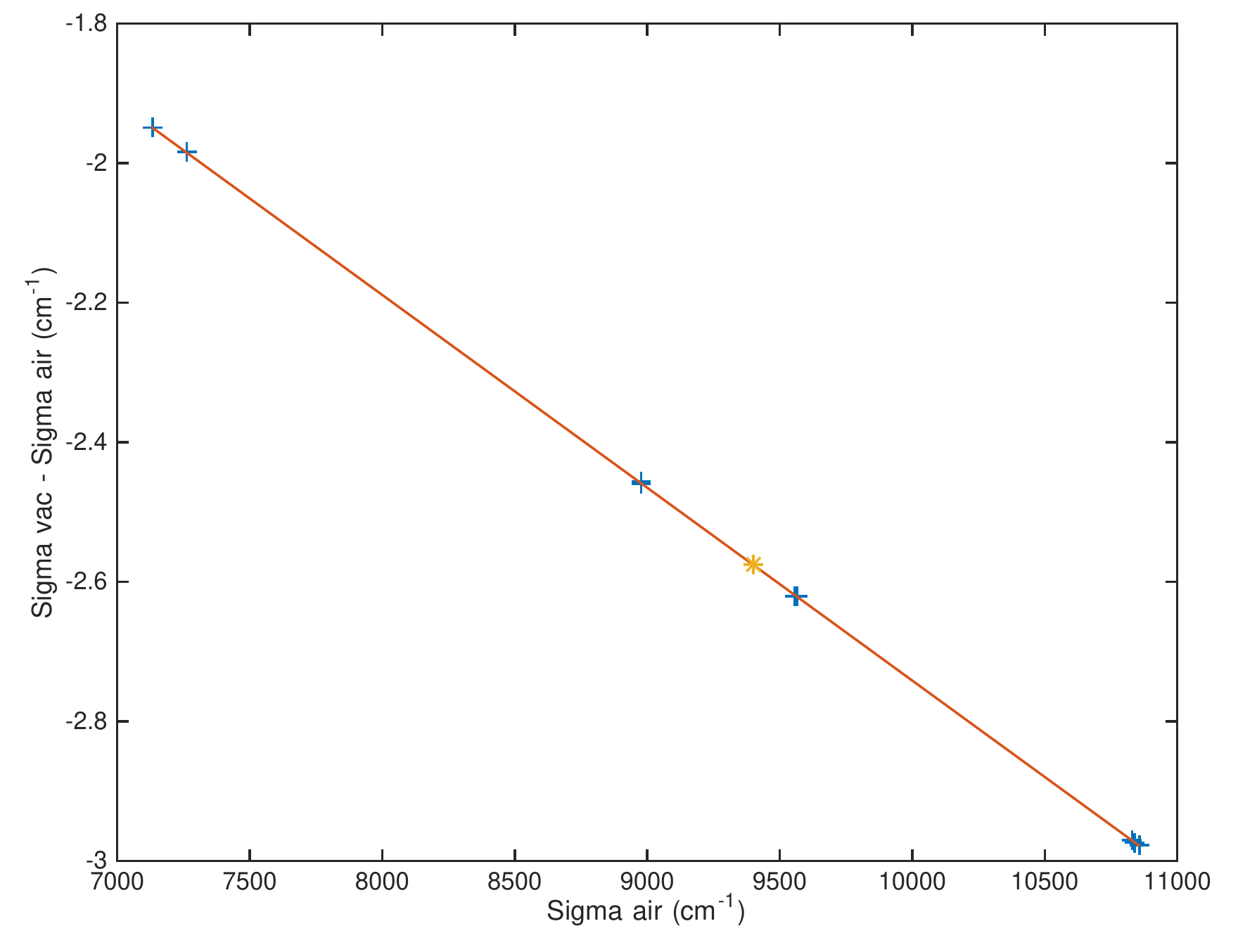}
\includegraphics[trim={0cm 0cm 0cm 0cm},clip,width=0.48\linewidth]{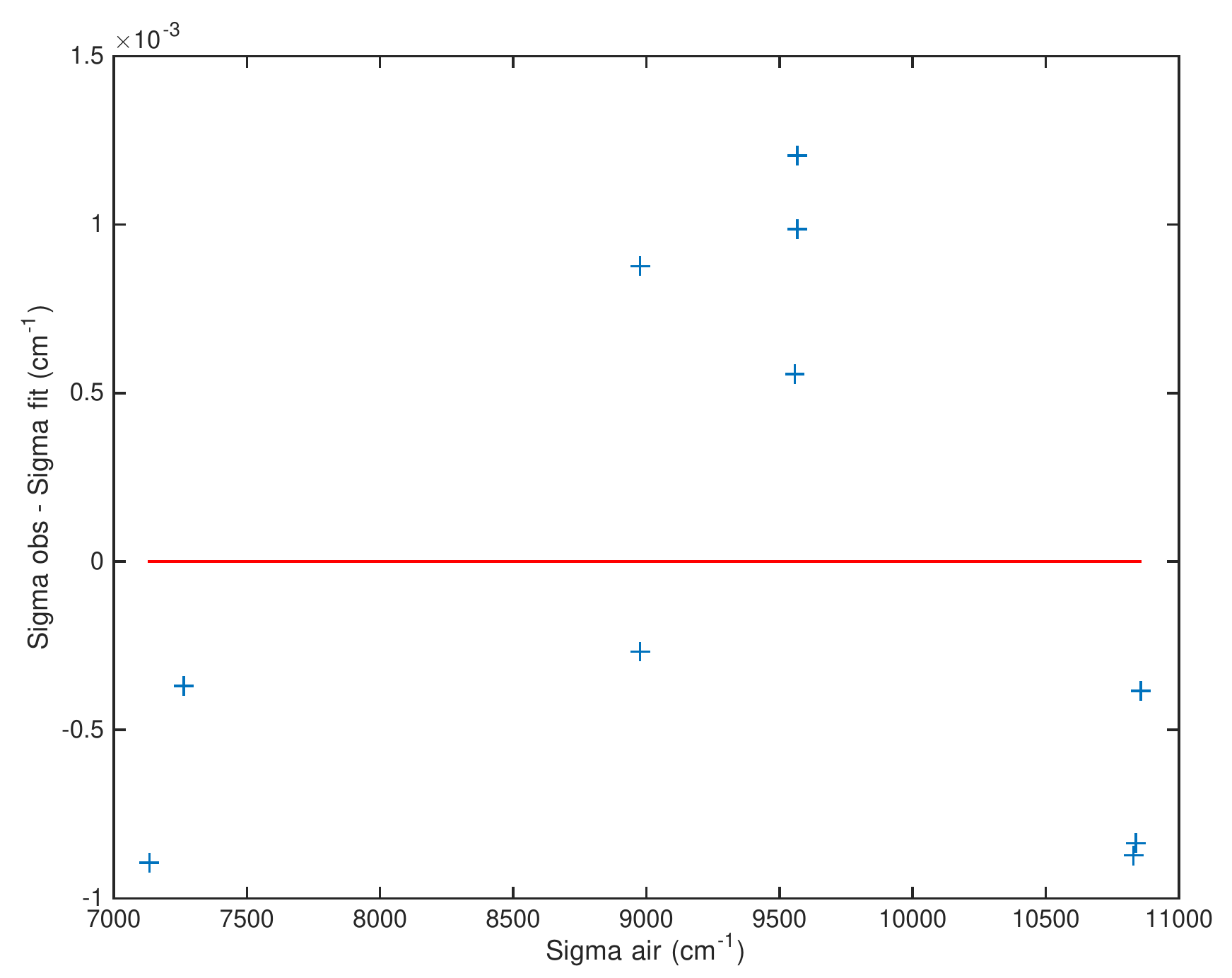}
\caption{Plots showing the correction to air wave numbers. {\it Left:} difference between air and vacuum wave numbers, and a linear fit to the existing values. {\it Right:} Difference between the reported wave numbers by \citet{jacobsson:67} and the values from the linear fit. The difference is of the order of 0.001 cm$^{-1}$ which is less than the stated uncertainty in the study by \citet{jacobsson:67}. }
\label{wavenumberfit}
\end{figure*}

\begin{figure*}[t]
\centering
\includegraphics[trim={0cm 0cm 0cm 0cm},clip,width=1.0\linewidth]{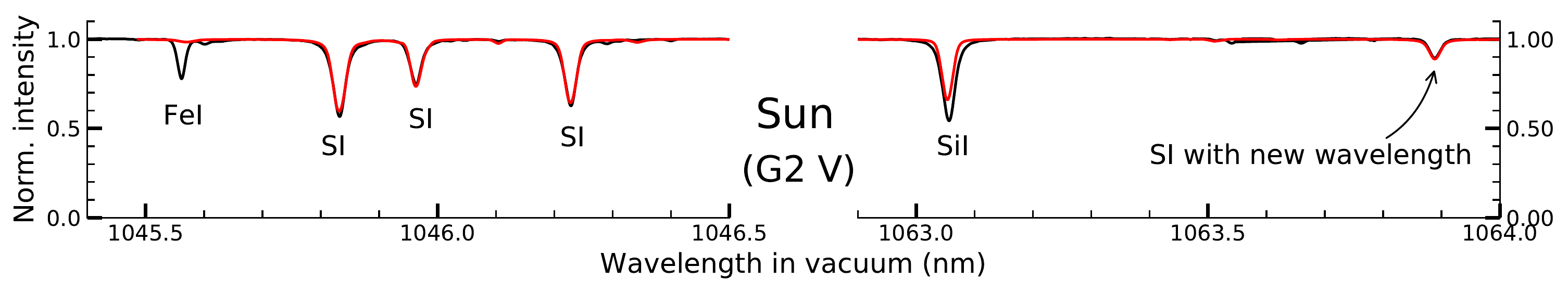}
\caption{Intensity spectrum of the solar center around the S\,{\sc i} triplet at 1045 nm and the moved S\,{\sc i} line. No further adjustments to the line strengths in the line lists from VALD have been made. The solar abundances \citep{solar:sme} are assumed and the $\log gf$ values of the S\,{\sc i} lines are those as listed in NIST, i.e. \citet{zerne:97} and \citet{gf_SI}. The Fe\,{\sc i} line and the Si\,{\sc i} lines obviously have wrong $\log gf$ values and can be adjusted to solar astrophysical values if wanted.}
\label{synt}
\end{figure*}

\section{Discussion and conclusions}

The unidentified feature at $\lambda_\mathrm{vac}=1063.891$\,nm  ($\lambda_\mathrm{air}=1063.600$\,nm) resembles a single spectral line, broadened similarly as nearby spectral lines, especially in the four cooler stars (see Figure \ref{fig:giano_spectrum}). Due to the fact that the telluric spectra do not show appreciable features in this region and the fact that the feature is at the same stellar rest-wavelength in all stars, the possibility of it being a telluric line is excluded and it therefore has to be a stellar feature. 

The feature becomes very shallow at temperatures below 5000~K and above 8000~K. This is a typical behaviour of a highly excited spectral line of a neutral ion; it disappears at high temperatures due to the ionization equilibrium and  at lower temperatures the line gets weaker due to lower degree of excitation.  However, no known line lies at this wavelength according to the NIST \citep{NIST:18} and VALD \citep{vald} databases.
The closest lines are much further away than any reasonable  uncertainty of measured laboratory wavelengths of any known atom, or the uncertainty of the wavelength scale of the observed spectra.

A first clue as to the  origin of the spectral feature can be found in the similar behaviour of the close-by sulphur-line triplet at 1046 nm, lines with an excitation energy of $\chi_\mathrm{exc}=6.9$\,eV. These lines originate from the $3s^23p^34s\,^3\mathrm S_1^o-3s^23p^34p\,^3\mathrm P_{0,1,2}$ transitions, have been measured in the laboratory by \citet{zerne:97}, and were successfully used in a work on the galactic chemical evolution of sulphur in the galaxy \citep[e.g.][]{caffau:07,jonsson:11}. They change in strength in a similar way as the feature at $\lambda_\mathrm{vac}=1063.891$\,nm, see Figure \ref{fig:giano_spectrum}. We have therefore searched for measured S\,{\sc i} lines in the literature. 

In the NIST database there is indeed an S\,{\sc i} line at $\lambda_\mathrm{vac}=1063.599$ ($\lambda_\mathrm{air}=1063.308$). This line has only slightly higher excitation energy than the triplet at 1046 nm, namely $\chi_\mathrm{exc}=8.6$\,eV and would fit in nicely with the line strength variation with effective temperatures of the observed stars.  The wavelength of the line is, however, too far away for being  accommodated within a random uncertainty in the wavelength measurement.

The relevant laboratory data for \ion{S}{i}, on which the NIST spectral data are derived,  is from the work by \citet{jacobsson:67}. They used a high-frequency discharge with sulfur dioxide as a light source to produce the sulfur lines. Since the wavelength region studied was the 'extraphotographic infrared', the light was fed into a 1 m scanning grating spectrograph operated with a nitrogen-cooled PbS-detector which fed a pen recorder producing the spectrum.
Their primary data are the air wavenumbers, $\sigma_{air}$, subsequently converted to air wavelengths, $\lambda_{air}$, and vacuum wavenumbers, $\sigma_{vac}$, in their Table 1. As can be seen there, for the 4s' $^1$D$_2$ - 4p' $^1$F$_3$  line the vacuum wavenumber, denoted by $\sigma_{obs}$, was not corrected for the refractive index to provide $\lambda_{air}$ = 1063.5993\,nm ($\sigma_{air} = 9402.037$), while the correction has been properly applied to all the other lines.
The correction amounts to about 2.6 cm$^{-1}$, which can fully account for the observed wavelength mismatch.

Unfortunately, neither the procedure for the conversion to vacuum data, nor the value used for the refractive index $n$ is presented in the paper. Since $n$ is slowly varying with wavenumber, we use the laboratory data of the neighbouring lines to interpolate the true laboratory data for the missing line, as presented in Figure \ref{wavenumberfit}. We find a wavenumber shift ($\sigma_\mathrm{vac}-\sigma_\mathrm{air}$) of $-2.576$\,cm$^{-1}$. We thus derive a new value of $\sigma_\mathrm{vac}=9399.461$\,\cm  ($\lambda_{vac}$=1063.8908\,nm). 

As is seen in the right panel of Figure \ref{wavenumberfit}, the difference between the laboratory value and the fitted value is around 0.001 cm$^{-1}$ or less, corresponding to 1 m\AA. This is an order of magnitude smaller compared to the stated uncertainty of the laboratory data, well justifying the linear interpolation. For the discussed lines, measured by \citet{jacobsson:67} by a photoelectric setup, they estimate the wavenumber accuracy to be better than 0.01 cm$^{-1}$, corresponding to 10 m\AA\ or 0.001~nm at these wavelengths. We estimate the uncertainty in our new values to be of the same magnitude. In addition, we choose to report the same number of digits for the numbers as the original reference. The new atomic wavelengths and wavenumber are reported in Table \ref{line}, along with the oscillator strength as listed in the NIST database \citep{NIST:18}, from the calculations by \citet{gf_SI}. The uncertainty of this value, quoted by the authors, is 2\%.

The energy of the $^1$F$_3$ level (upper level of the transition) is, due to the sparse number of lines, derived only from one line (i.e. the one under investigation in this paper). Since the current value is based on the the wrong wavenumber, we thus use the new wavenumber to derive a corrected energy of the 3s$^2$3p$^3$($^2$D)4p $^1$F$_3$ level. The new excitation energy  of the level is $E_\mathrm{exc}$  =  78637.303 cm$^{-1}$   = 9.7497830(10) eV

\begin{table*}
\caption{New atomic line data for the S\,{\sc i} line (3p$^3$($^2$D)4s $^1$D$_2$ - 3p$^3$($^2$D)4p $^1$F$_3$).  Uncertainties are given in brackets.}             
\label{line}      
\centering          
\begin{tabular}{l c }     
\hline\hline 
Spices & \ion{S}{I}\\
Wavelength (in vacuum) &  1063.8908(11) nm \\
Wavelength (in air) &  1063.5993(11) nm \\
Wavenumber & 9399.461(10) cm$^{-1}$\\
$\log gf$ \citep[NIST:][]{gf_SI}& 0.391(10)\\
$E_\mathrm{exc}$ (lower) \citep{jacobsson:67} & 8.5844037(10) eV\\ 
$E_\mathrm{exc}$ (upper)   & 78637.303(10) cm$^{-1}$ \\
$E_\mathrm{exc}$ (upper)   & 9.7497830(10) eV \\
Lower level Term & $3s^23p^3(^2\rm D^o)\,4s\,^1\rm D^o_2$\\
Higher level Term & $3s^23p^3(^2\rm D^o)\,4p\,^1\rm F_3$\\
\hline                  
\end{tabular}
\end{table*}

Being confident that the published wavelength of the S\,{\sc i} line from the original reference \citep{jacobsson:67} is wrong, due to an air to vacuum conversion mistake, we finally check our new wavelength by synthesizing the solar-center intensity-spectrum 
observed with the Fourier Transform Spectrometer at the McMath/Pierce Solar Telescope at Kitt Peak \citep{solar_atlas_0.9-1.1um} for the wavelength region around the S\,{\sc i} line. To do this we use the spectral synthesis code {\it Spectroscopy Made Easy, SME} \citep{sme,sme_code}, in which the radiative transfer and line formation is calculated for a model atmosphere defined by the fundamental stellar parameters of the Sun. 
{\it SME} uses a grid of model atmospheres in which the code interpolates for a given set of fundamental parameters of the analyzed star. We use 1-dimensional MARCS models, which are hydrostatic model photospheres in  plane-parallel geometry, computed assuming LTE, chemical equilibrium, homogeneity, and conservation of the total flux (radiative plus convective, the convective flux being computed using the mixing-length recipe) \citep{marcs:08}. We calculate the synthetic spectrum of the solar spectrum as an intensity spectrum of the solar center.

In Figure \ref{synt} we show our synthesised solar spectra together with the observed intensity spectrum of the solar center. As can be seen the previously unidentified line fits well and in tandem with the sulfur triplet at 1046 nm. We can thus confidently move the S\,{\sc i} line in the NIST database to its correct wavelength as determined from the GIANO-B spectra. The final parameters of this S\,{\sc i} line are summarized in Table \ref{line}.

\begin{acknowledgements}
This research has been partly supported by the Royal Physiographic Society in Lund through Stiftelsen Walter Gyllenbergs fond and M{\"a}rta och Erik Holmbergs donation. 
Data is used from the Fourier Transform Spectrometer at the McMath/Pierce Solar Telescope
situated on Kitt Peak, Arizona,  operated by the National Solar
Observatory, a Division of the National Optical Astronomy Observatories.
NOAO is administered by the Association of Universities for Research in
Astronomy, Inc., under cooperative agreement with the National Science
Foundation.
\end{acknowledgements}

\bibliographystyle{aa}

\end{document}